\documentclass[twocolumn,showpacs,preprintnumbers,amsmath,amssymb,nofootinbib]{revtex4}

\usepackage{graphicx}% Include figure files
\usepackage{dcolumn}% Align table columns on decimal point
\usepackage{bm}% bold math
\usepackage{epsf}
\begin{document}

%\preprint{APS/123-QED}

\title{Time Series with Tailored Nonlinearities}

\author{C. R\"ath$^1$, I. Laut$^1$}
 \affiliation{ $^1$ Deutsches Zentrum f\"ur Luft- und Raumfahrt, Forschungsgruppe Komplexe Plasmen, M{\"u}nchner Stra{\ss}e 20, 82234 We{\ss}ling, Germany
        }

\date{\today}% It is always \today, today,

\begin{abstract}
It is demonstrated how to generate time series with tailored nonlinearities by inducing well-defined constraints on the Fourier phases.
Correlations between the phase information of adjacent phases and (static and dynamic) measures of nonlinearities are established and their origin is explained.
By applying a set of simple constraints on the phases of an originally linear and uncorrelated Gaussian time series, 
the observed scaling behavior of the intensity distribution of empirical time series can be reproduced.
The power law character of the intensity distributions being typical for e.g. turbulence and financial data 
can thus be explained in terms of phase correlations. %as the most general way to define nonlinearities. 
\end{abstract}



\maketitle

{\it Introduction.} 
The clearest yet  most general definition of nonlinearity in time series $g(t)$ is given in the 
Fourier representation 
\begin{equation}
G(k) = FT(g(t)) = \frac{1}{N} \sum_{t=0}^{N-1} g(t) e^{-i2\pi kt / N} 
\end{equation}
of the data.
Linear time series are fully characterized by the modulus $|G(k)|$ of the complex valued Fourier coefficients $G(k) = |G(k)| e^{i \phi(k)} $, while 
the phases $\phi(k)$ are uncorrelated and uniformly distributed in the interval $\phi \in [-\pi;\pi]$.
{\it Any} nonlinearity is coded in the Fourier phases $\phi(k)$ and correlations among them.  
Deviation from the randomness of the phases are thus equivalent to the presence of nonlinearities in the time series.
As yet, only little attention has been paid so far to the explicit analysis of the information contained 
in the Fourier phases to characterize nonlinearities, 
although a lot of insights about nonlinearities may be gained by better understanding the meaning of phases. \\
The definition of nonlinearity via the randomness of Fourier phases is -- on the other hand -- at the heart of 
algorithms for generating so-called surrogate data sets, which were developed 
to test for weak nonlinearities in a model-independent way \cite{Theiler92}. These surrogates are supposed to have the same 
linear properties as a given data set, while all nonlinearities are wiped out. The removal of the nonlinear correlations is achieved by 
replacing the phases $\phi(k)$ with a set of uncorrelated and uniformly distributed ones.
Refinements of the Fourier-based methods for generating surrogates aimed at preserving both the power spectrum $|G(k)|^2$ and the 
amplitude distribution of the time series $g(t)$ in real space  \cite{Theiler92, Schreiber96,Keylock06,Keylock10}. 
The addition of (iterative) rank-ordered remapping of the phase randomized 
data onto the original amplitude distribution led to surrogates with the desired amplitude spectrum \cite{Theiler92, Schreiber96}.
Applying the IAAFT method in the wavelet domain allowed for the generation of surrogates which also preserve the local mean and variance of the original signal \cite{Keylock06,Keylock10}.\\   
However, it was found recently that these (iterated) amplitude adjusted ( (I)AAFT ) surrogates  
may not be linear, since the randomness of the phases is guaranteed only 
before the first remapping step. One can rather find phase correlations in surrogate realizations that may result in a non-detection 
of nonlinearities in time series \cite{Raeth12}. But this obvious flaw of (I)AAFT surrogates became a virtue as 
significant correlations between phase statistics and a measure for nonlinearity were found for the first 
time (see \cite{Raeth12} and insets in Fig. \ref{fig:noise_nlpe}) .\\  
Connections between correlations among Fourier phases and higher order statistics could also identified 
by analyzing the cosmic microwave background radiation (CMB). Several studies of both the WMAP and PLANCK data 
involving surrogates revealed that there are phase correlations at large scales in the CMB which lead to pronounced 
anisotropies (see e.g. \cite{Raeth09,Raeth11,Ade13}). 
Recently it was demonstrated that the observed phase correlations can gradually be diminished when subtracting 
suitable best-fit (Bianchi-)template maps. The weaker phase correlations lead in turn to a vanishing signature of anisotropy 
as identified with higher order statistics \cite{Modest14}. \\%, i.e. with scaling indices and Minkowski functionals \cite{Modest14}.
The relations between phase information and higher order statistics in (I)AAFT surrogates and the CMB data 
were only found in a heuristic manner.
%Although in both cases relations between phase information and higher order statistics could be established, 
%the drawback is that the relations were found in a heuristic manner.
%%%%%%%%%%%%%%%%%%%%%%%%%%%%%%%%%
%%%%%%%%%%%%%%%%%%%%%%%%%%%%%%%%%
%Fig1%%%%%%%%%%%%%%%%%%%%%%%%%%%%%%%
\begin{figure}[h]
\centering
 %\vspace{1cm}
  \includegraphics[width=\columnwidth,angle=0]{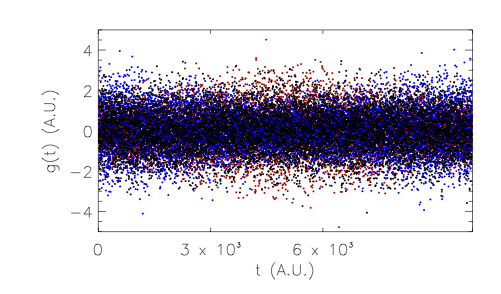}
   \caption{Gaussian random uncorrelated noise (black). The colored points show time series with linear phase 
   correlations among adjacent phases ($\Delta = 1$) with $\sigma_{\eta}=1.0$ and  $d\phi =0$ (blue) and  $d\phi =\pi$ (red). 
    %For more details see text.  
    \label{figure20}}
\end{figure}
%%%%%%%%%%%%%%%%%%%%%%%%%%%%%%%%
 % Fig2 %%%%%%%%%%%%%%%%%%%%%%%%%%%%%%%%%%%%
\begin{figure*}%%%%%%%%%%%%%%%%%%%%%%%%%%%%%%%%%%%%%%%%
%\begin{figure}%%%%%%%%%%%%%%%%%%%%%%%%%%%%%%%%%%%%%%%%

%\centering
%\includegraphics[width=\columnwidth]{noise_nlpe_vs_c_Tau250_d3}
%\includegraphics[width=\columnwidth]{noise_deg_vs_c_Tau250_d3}
%\includegraphics[width=\columnwidth]{mkn_nlpe_vs_c_Tau250_d3}
%\includegraphics[width=\columnwidth]{mkn_deg_vs_c_Tau250_d3}

\includegraphics[width=8cm]{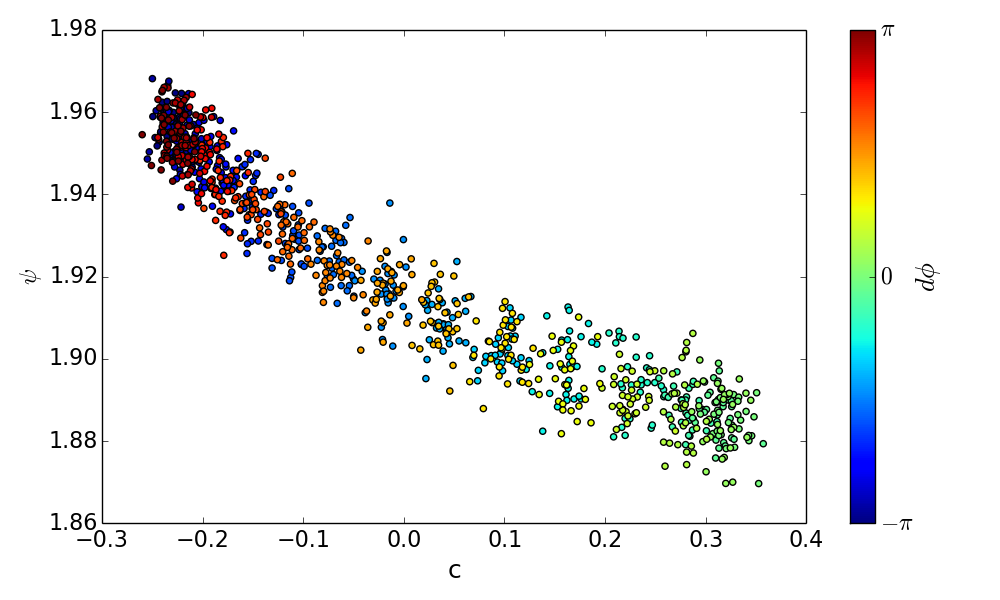}
\includegraphics[width=8cm]{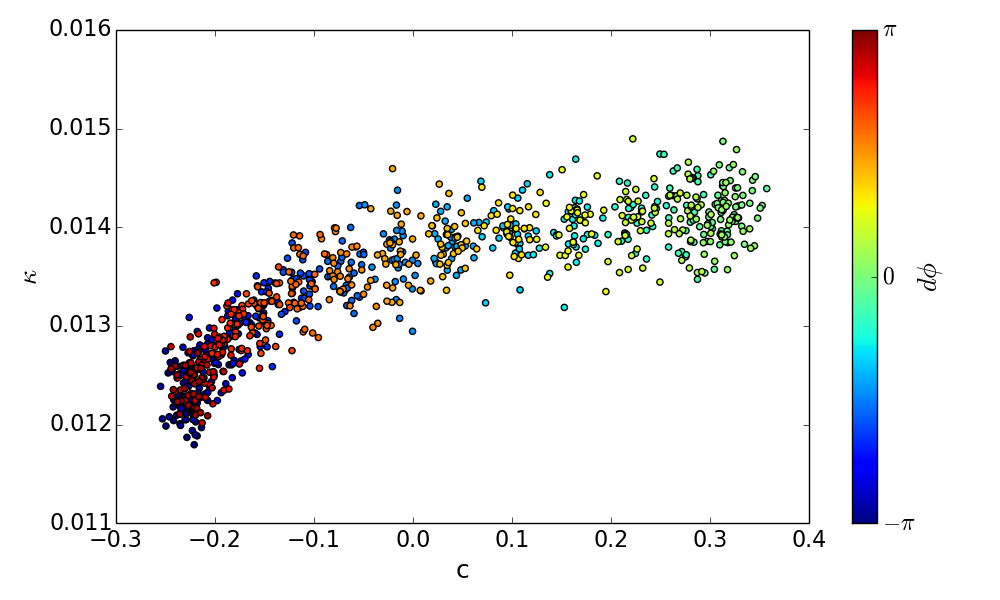}
\includegraphics[width=8cm]{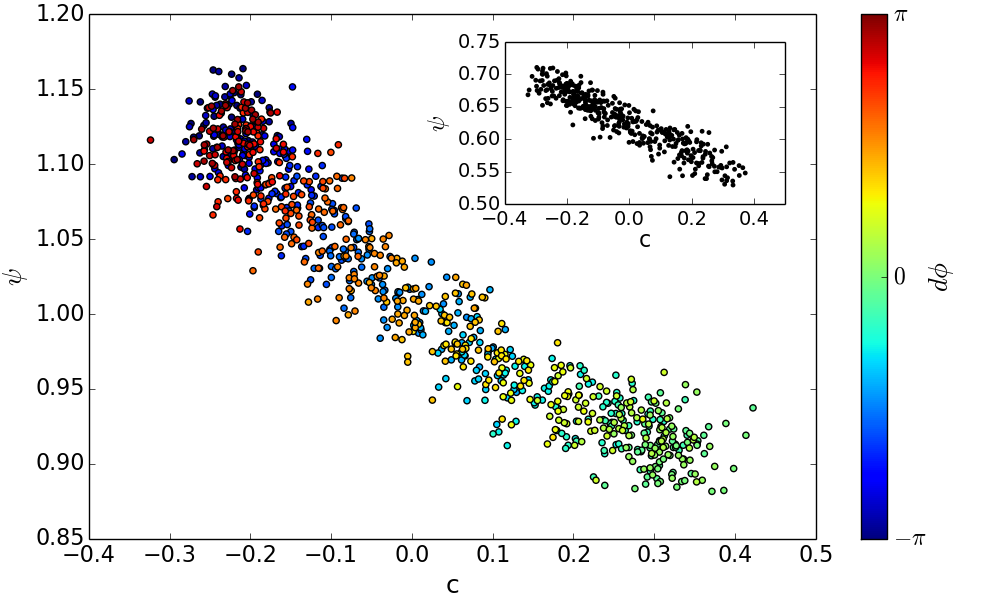}
\includegraphics[width=8cm]{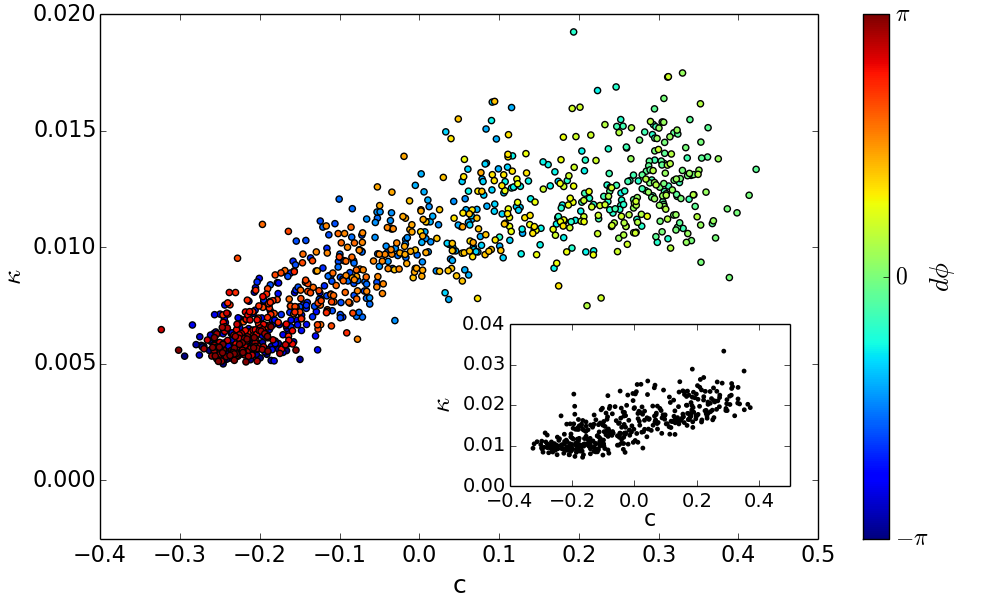}

\caption{Upper Row: Nonlinear prediction error $\psi$  (upper left) and  average connectivity $\kappa$ (upper right) 
               versus phase correlation coefficient $c(\Delta =1)$ for 900 time series with imposed phase correlations derived from Gaussian noise as input time series $g(t)$.
                Lower row: Same as upper row but with the Mrk 766 X-ray observation as input time series $g(t)$.  The data from 
                revolution 999 was binned with a binsize of 50 s leading to a time series with 1540 points. The insets show the corresponding results for 400 IAAFT surrogates.}
\label{fig:noise_nlpe}
%\end{figure}%%%%%%%%%%%%%%%%%%%%%%%%%%%%%%%%%%%%%%%%
\end{figure*}%%%%%%%%%%%%%%%%%%%%%%%%%%%%%%%%%%%%%%%%

To allow for a systematic investigation of phase correlations and their corresponding nonlinearities in time series, it is desirable 
to start with the phases, constrain their correlations in a tunable and reproducible way and study the effects on the nonlinear statistics.\\
Here, we present a method to generate time series with such tailored nonlinearities by imposing well-defined  
correlations on the Fourier phases and demonstrate how deviations from linearity can be understood in terms of phase information.\\
{\it Methods.} To address the relationship between phase correlations and measures for nonlinearity
we calculate the nonlinear prediction error (NPLE) \cite{Sugihara90} as an example for a dynamical 
complexity measure with a good overall performance \cite{Schreiber97} 
and the average connectivity of (recurrence) networks as an example for a structural 
complexity measure \cite{Donner10,Laut14}. The calculation of both measures
relies on the representation of the time series in an artificial phase space, which is obtained
using the method of delay coordinates \cite{Packard80}. 
This is accomplished by using time delayed versions of the observed time series as
coordinates for the embedding space. The multivariate vectors in the $d$-dimensional 
space are expressed by $\vec{g}_t= ( g_t,  g_{t + \tau}, g_{t + 2 \tau},\ldots ,g_{t+(d-1) \tau}) \;,$
where $\tau$ is the delay time and $g_t$ denotes the value  of the (discretized) time series at time step $t$.\\
The comparison of the predicted behavior of the embedded time series based on the local neighbors
with the real trajectory of the system leads to the definition of the NLPE $\psi$ as
\begin{eqnarray}
\psi & = & \psi(d,\tau,T,N) \\ \nonumber
           &= &\frac{1}{(M-T-(d-1)\tau)} \sqrt{ \sum_{n=(d-1)\tau}^{M-1-T} [\vec{g}_{t+T} -F(\vec{g}_t)]^2}   \;,   
\end{eqnarray}
where $F$ is a locally constant predictor, $M$ is the length of the time series, 
and $T$ is the lead time. The predictor $F$ is calculated by averaging over future values 
of the $N=d+1$ nearest neighbors in the delay coordinate representation.
We found that $\psi$ remains rather constant for $T > 5$, 
thus a value of $T=5$ was used for this study.
The dimension of the embedding space $d$ 
and the delay time $\tau$ have to be set appropriately. 
Since the time series of an Active Galactic Nuclei (AGN) being studied in the following consists of less than 1600 data points, we use a low embedding dimension $d=3$. 
Due to the long correlation time of this time series, we chose a relatively large delay time $\tau = 250$  according to the criterion of zero crossing of the 
autocorrelation function \cite{Fraser86}. 
To allow for a direct comparison, we use the same values $d=3$ and $\tau = 250$ for the other time series with imposed phase correlations. 
The structural complexity of a time series with a limited number of points can be characterized with recurrence networks \cite{Marwan09}.
They are based on  recurrence plots \cite{Eckmann87}, which describe how often pairs 
of points of a time series in the embedding space representation
come close to each other.
Linking such nearby points in a network representation of the data and omitting self-loops leads to the definition of the
adjacency matrix $A_{i,j}$ of the recurrence network \citep{Donner10} 
\begin{equation}
 \label{equ_ts_Aij}
 A_{i,j} = \Theta\left(\epsilon - \| \mathbf{g}_i - \mathbf{g}_j  \| \right)- \delta_{i,j},
\end{equation}
where the $\{\mathbf{g}_n\}$ are the data points in embedding space and $\epsilon$ is an appropriate threshold.
$A_{i,j}$ contains the whole information about the network.
A common measure for the topological structure of the network is the 
average connectivity $\kappa$ which is calculated by 
\begin{equation}
 \label{equ_network_average_connectivity}
 \kappa = \frac{1}{n(n-1)} \sum_{\nu=0}^{n-1} k_{\nu},
\end{equation}
where $k_\nu =  \sum_{i=0}^{n-1} A_{\nu, i}$ is the degree of node $\nu$. 
If the attractor of the nonlinear system is reconstructed with appropriate embedding parameters this network measure can be used as a test for nonlinearity. 
The threshold $\epsilon$ is chosen such that $\kappa = 0.01$ for the original time series. 
The same threshold is then used for the Gaussian time series with imposed phase correlations.\\
To get a visual impression of correlations among the Fourier phases it is convenient to make use of so-called phase maps \cite{Chiang02}. 
A  phase map is defined as a two-dimensional set of points $G=\{ \phi_k,\phi_{k+\Delta} \}$ where $\phi_k$ is the 
phase of the $k^{th}$ mode of the Fourier transform and $\Delta$ a frequency  delay. 
To quantify the degree of correlation between the phases $\phi$ and $\phi + \Delta$ we 
calculate the correlation coefficient
$c(\Delta)$,
\begin{equation}
  c(\Delta) = \frac{\langle \phi(k) \phi(k+\Delta) \rangle}{\sigma_{\phi(k)} \sigma_{\phi(k+\Delta)} } \;.
\end{equation}
Note that by using $c(\Delta)$ as correlation measure we restrict ourselves to the simplest way 
of quantifying correlations among the phases that is only sensitive to linear correlations.

{\it Time series with phase correlations.} As outlined in \cite{Raeth11}, (I)AAFT surrogates 
can contain phase correlations leading to statistically significant  high or low 
values of $c(\Delta)$. A closer look at the corresponding phase maps reveals that 
the (anti-)correlations originated from stripe-like, patterns along the diagonal (i.e. with slope of one) 
or shifted relative to it. These patters thus indicate that phase pairs are linearly correlated with each other. 
One can further notice that for the time series stemming from X-ray observations of 
the AGN Mrk 766 the phase correlations are most pronounced 
for $\Delta=1$. We reproduce such signatures by imposing correlations in the phase distribution in the following way:
The values for the phases $\phi(k)$ are iteratively determined by relating $\phi(k+\Delta)$ with $\phi(k)$ by
\begin{equation}
 \label{equ_phase_corr}
 \phi(k+\Delta) = \phi(k) + d\phi + \eta
\end{equation}
with $d\phi$ being a shift constant ranging from $- \pi$ to $\pi$ and $\eta$ describing 
a (Gaussian) noise term with given standard deviation $\sigma_{\eta}$. 
In the phase map picture $\eta$ controls the width of the stripes and $d\phi$ defines its position.
The iteration is performed over the frequencies $k$, where $k_s$ denotes the starting value and 
$dk$ the step size of the iteration.  
$\phi(k_s)$ is drawn from a uniform distribution within the interval $[-\pi, \pi]$. The same is true for $\phi(k)$ if this
phase has not been set in a previous iteration step. 
In our first example we are interested in only correlating adjacent phases. Thus we apply Eq.  \ref{equ_phase_corr}
with $\Delta=1$ to a Gaussian time series with zero mean and standard deviation of one. 
The step size is chosen to $dk= 2$. Thereby every phase is correlated to exactly one other phase for $\Delta = 1$, 
while the phases are not correlated for any frequency delay  $\Delta$ greater than one.\\
Fig. \ref{figure20} shows how these phase correlations alter the time series. It becomes clearly visible that the correlations of adjacent phases
induce fluctuations of the variance. Specifically, one recognizes a time interval where the fluctuations are larger than for the noise and 
another region where the fluctuations are smaller. Note that the overall mean and standard deviation of the time series
are exactly preserved since the power spectrum is kept constant.
The shift constant controls the position of the region with higher fluctuations. If $d\phi=0$, this region  is located in the middle 
of the time series and it shifts towards the ends of the time series when  $d\phi$ approaches $\pm \pi$.
By testing different values of $\Delta$ we further found that the number of regions with high fluctuations is given by the value for $\Delta$.
In Fig. \ref{fig:noise_nlpe} we show the nonlinear prediction error  $\psi$ and the mean connectivity $\kappa$ as a function of 
the correlation coefficient $c(\Delta=1)$. The results are displayed for time series with imposed phase correlations (only) for $\Delta =1$  and varying $d\phi$ 
as derived from Gaussian noise and from the X-ray observation of the AGN Mrk 766.
One can see that the shift constant $d\phi$ controls the (anti-)correlations of the phases. 
More importantly, it becomes obvious that both the nonlinear prediction error and the mean connectivity are highly (anti-)correlated
with the phase correlations as measured with $c(\Delta)$. 
Knowing that $d\phi$  also controls the position of the regions with higher and lower fluctuations, we can now get a much more 
detailed understanding of how the phase correlations influence the calculation of the NLPE  and the average connectivity.
The embedding with the delay time of $\tau=250$ in three dimensions leads to a truncation of the last part of the time series. 
Depending on whether the remaining time series has lower ($d\phi \approx \pm \pi$) or larger ($d\phi \approx 0$) fluctuations, 
one obtains larger or lower values for the NLPE leading to the observed anti-correlation between psi and c.
Similarly, lower fluctuations in the time series lead to a more connected recurrence network and vice versa, correlating $\kappa$ and $c(\Delta)$.
%%%%%%%%%%%%%%%%%%%%%%%%%%%%%%%%%%%%%%
\begin{table} 
\caption{Parameters defining the imposed phase correlations \label{table1}}
\begin{ruledtabular}
\begin{tabular}{lccccc}
     &  $\Delta $ &  $dk$ & $k_s$ & $d\phi$ & $\eta$    \\ \hline
     1. Iteration & 1 & 2 & 1  & 3.1415 & 0.1 \\
     2. Iteration & 3 & 3 &  1 & 3.0        & 0.08\\
     3. Iteration & 3 & 3 &  2 & 3.0 &  0.3 \\
     4. Iteration & 5 & 5 &  2 & 1.4 & 0.2 \\
     5. Iteration & 7 & 7 & 3 & 3.1415 & 0.25 \\ 
     6. Iteration & 50 & 50 & 2& 3.0 & 0.1 \\
\end{tabular}
\end{ruledtabular}
\end{table}
%%%%%%%%%%%%%%%%%%%%%%%%%%%%%%%%

%%%%%%%%%%%%%%%%%%%%%%%%%%%%%%%%
\begin{figure}[h]
\centering
 %\vspace{1cm}
  \includegraphics[width=\columnwidth,angle=0]{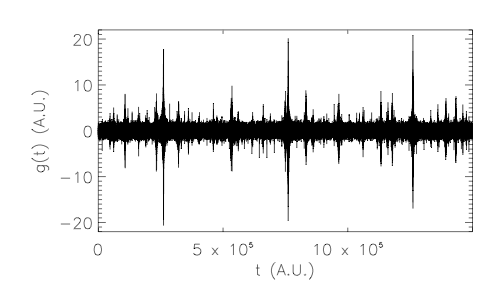}
   \caption{Time series $g(t)$ that is obtained from white Gaussian noise by imposing a  set of six different linear phase correlations.   \label{figure300}}
\end{figure}
%%%%%%%%%%%%%%%%%%%%%%%%%%%%%%%%

%%%%%%%%%%%%%%%%%%%%%%%%%%%%%%%%
\begin{figure}[h]
\centering
 %\vspace{1cm}
  \includegraphics[width=\columnwidth,angle=0]{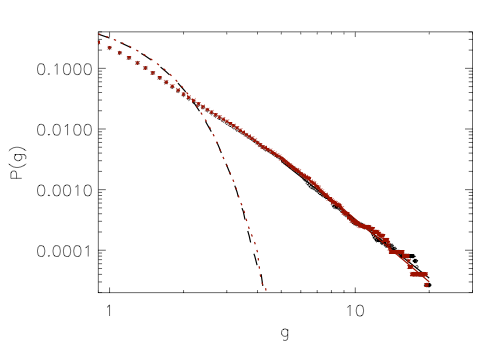}
     \caption{Cumulative probability distribution $P(g)$ of the normalized positive (black) and negative (red) values of $g(t)$. 
      The black dashed and red dotted lines show the respective distributions for the initial white Gaussian noise. \label{figure200}}
\end{figure}
%%%%%%%%%%%%%%%%%%%%%%%%%%%%%%%%

%%%%%%%%%%%%%%%%%%%%%%%%%%%%%%%%
\begin{figure}[h]
\centering
 %\vspace{1cm}
  \includegraphics[width=\columnwidth,angle=0]{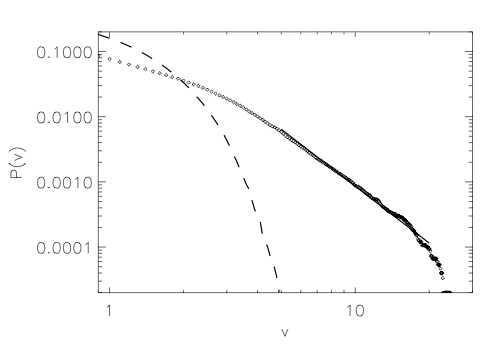}
   \caption{Cumulative probability distribution $P(v)$ of the normalized volatility $v$. 
   The black dashed line shows the respective distributions for the initial white Gaussian noise. \label{figure100}}
\end{figure}
%%%%%%%%%%%%%%%%%%%%%%%%%%%%%%%%

In a second example we extend the formalism to generate nonlinear time series with well-defined nonlinearities by simultaneously imposing
linear phase correlations for a set of different frequency  delays $\Delta$. This is achieved by  iteratively applying Eq. \ref{equ_phase_corr}
starting with low values of $\Delta$ and then proceeding to higher ones. 
Iterating over increasing frequency  delays $\Delta$ ensures that phase correlations that were imposed in previous iteration steps 
are at least in parts preserved when new constraints for phase correlations at larger $\Delta$ are added.
Table \ref{table1} summarizes the parameters for the six iterations used in our example.  Fig.  \ref{figure300} shows the time series which 
is obtained when the six constraints on phase correlations are imposed on white Gaussian noise. 
One has to note that the time series has no linear correlation as the modulus $|G(k)|$ of the 
Fourier transform of the original random time series is left untouched.
In the time series with phase correlations one can clearly identify a number 
of time intervals with larger fluctuations whose number, position and  strength 
are controlled by the parameters $\Delta$, $dk$ and $\eta$, respectively. 
The statistical properties of such a nonlinear time series can thus be tailored in a refined manner.  
The time series in this example was generated such that it resembles data often observed in economic 
time series \cite{Mantegna07}, where especially data from stock indices  
show intermittent behavior, i.e. extreme events, patterns of volatility clustering and phase correlations, while the autocorrelation vanishes.
The distribution of the fluctuations is further analyzed by calculating the cumulative probability 
distribution $P(g)$ of the normalized positive and negative values of $g(t)$ (see Fig. \ref{figure200}).
We find the expected leptokurtic distribution whose tail can be fitted with a power law $P(g) \sim g^{- \alpha}$ with 
$\alpha = 3.25 \pm 0.16$ for the positive tail and $\alpha = 3.37 \pm 0.47$ for the negative tail in the region $5 \le g \le 20$.
These numbers are in remarkable agreement  with those obtained from empirical studies of market indices \cite{Gopikrishnan99}.
We further studied the statistical properties of the volatility $v(t)$ as defined as the average of $ |g(t)|$ over a time window of length $N$, i.e.
$v(t) = 1/N \sum_{t'= t}^{t+N-1} |g(t')|  $. 
Fig. \ref{figure100} shows the cumulative probability distribution $P(v)$ for $N=5$. As expected we find a distribution with fat tails, which can
be fitted by $P(v) \sim v^{- \beta}$ with $\beta = 2.93 \pm 0.49$ in the region  $5 \le v \le 20$. 
Again, this is in very good agreement with the scaling properties 
of the volatility of price fluctuations observed in empirical data \cite{Liu99}.\\
Finally, we note that analogies between price dynamics of market indices 
and the velocity differences in three-dimensional fully developed turbulence 
have been pointed out by several authors (see e.g. \cite{Ghashghaie96, Mantegna97}).   
Consequently, the fat tails in the probability density functions of turbulence data may be also be understood in terms 
of phase correlations allowing for a better characterization and discrimination of different 
scenarios of turbulence.

{\it Summary.} We have presented a new method to generate time series 
with well-defined nonlinearities by imposing linear correlations among the Fourier phases.
We have shown that the phase correlations between adjacent phases are tightly 
related with higher order statistics being estimated for the time series. 
These "Wiener-Khinchin-like" connections between phase information and higher order statistics 
are to a large extent independent of the input time series.
Furthermore, the scaling of fluctuation and of the volatility of a time series can be understood in terms of a set of linear phase correlations. 
We expect that further studies with time series and also spatial structures with tailored nonlinearities, for which not only linear but more complex 
constraints on the Fourier phases are imposed, will shed more light  on both the meaning of Fourier phases and the different kinds of 
nonlinearities as they are observed in nature. 

{\it Acknowledgments.} This work has made use of observations obtained with {\it XMM-Newton}, an ESA science mission
with instruments and contributions directly funded by ESA member states and the US (NASA).

\bibliography{ts_with_tailored_nonlinearities}

\end{document}